\documentclass[aps,pre,twocolumn,groupedaddress]{revtex4-1}

\usepackage{euscript,amsmath,amssymb,amsfonts,graphicx,bm,amsthm,mathtools,bbm}

  \usepackage{paralist}
  \usepackage{epstopdf}
  \usepackage{graphics} 
 \usepackage[latin1]{inputenc}
\usepackage[caption=false]{subfig}
\usepackage{soul}

\usepackage{latexsym,amssymb,enumerate,amsmath,amsthm} 
  \usepackage{graphicx} 
  \usepackage{tikz}
  \usepackage{hyperref}
\usepackage{latexsym,amssymb,enumerate,amsmath} 
\usepackage{soul,xcolor}

\newcommand{\dd}{\textup{d}}

\def\E{\mathbb{E}}
\def\P{\mathbb{P}}

\def\<{\langle}
\def\>{\rangle}
\def\PP{\mathcal{P}}

\newtheorem{theorem}{Theorem}

\theoremstyle{plain}
\theoremstyle{remark}

\theoremstyle{definition}

\begin{document}


\title[]{Cover times of many random walkers on a discrete network}



\author{Hyunjoong Kim}
\affiliation{Department of Mathematics,
University of Houston, Houston, TX 77204, USA}

\author{Sean D. Lawley}
\email[]{lawley@math.utah.edu}
\affiliation{Department of Mathematics, University of Utah, Salt Lake City, UT 84112 USA}


\date{\today}

\begin{abstract}
The speed of an exhaustive search can be measured by a cover time, which is defined as the time it takes a random searcher to visit every state in some target set. Cover times have been studied in both the physics and probability literatures, with most prior works focusing on a single searcher. In this paper, we prove an explicit formula for all the moments of the cover time for many searchers on an arbitrary discrete network. Our results show that such cover times depend only on properties of the network along the shortest paths to the most distant parts of the target. This mere local dependence contrasts with the well-known result that cover times for single searchers depend on global properties of the network. We illustrate our rigorous results by stochastic simulations. 
\end{abstract}


\maketitle


\section{Introduction}

Cover times measure the speed of exhaustive search and have been studied in both the physics literature \cite{nemirovsky1990, yokoi1990, brummelhuis1991, hemmer1998, nascimento2001, zlatanov2009, mendoncca2011, chupeau2014, chupeau2015, majumdar2016, grassberger2017, cheng2018, maziya2020, dong2023, han2023} and the probability literature \cite{aldous1983, aldous1989, broder1989, aldous1989b, kahn1989, dembo2003, dembo2004, ding2012, belius2013, belius2017}. A cover time is defined as the time it takes a random searcher to visit every state in some target set of states (the target set is often taken to be the entire state space). Cover times characterize the timescales in a number of disparate applications \cite{chupeau2015}, including computer and internet search algorithms, animals collecting food and other resources, the immune system hunting pathogens (e.g.\ viruses, bacteria), and robots cleaning an area or combing for dangers (e.g.\ mines, explosives, chemical leaks).

To describe precisely, let $X=\{X(t)\}_{t\ge0}$ denote the path of a searcher (i.e.\ a random walk) on a discrete state space $I$ (i.e.\ a network of nodes). Let $S=\{S(t)\}_{t\ge0}$ denote the set of nodes visited by the searcher by time $t\ge0$,
\begin{align*}
    S(t)
    :=\cup_{s=0}^t X(s).
\end{align*}
The cover time of some target set of nodes $U_{\text{T}}\subseteq I$ is the time it takes the searcher to visit every node in $U_{\text{T}}$,
\begin{align*}
    \sigma
    :=\inf\{t>0:U_{\text{T}}\subseteq S(t)\}.
\end{align*}

Cover times can be contrasted with an alternative measure of search time called a first passage time (FPT) \cite{redner2001}. FPTs measure the time it takes a searcher to find a single target state (possibly out of a set of target states) and are defined mathematically as
\begin{align*}
\tau
:=\inf\{t>0:X(t)\in U_{\text{T}}\}.
\end{align*}
While FPTs have been studied more extensively than cover times, cover times are the important observable in any scenario requiring the discovery of multiple targets.

Analytical studies of cover times have mostly been carried out in the limit of a large network \cite{aldous1989, nemirovsky1990, brummelhuis1991, dembo2004, ding2012, belius2013, chupeau2015}. In the case of a Markovian, non-compact random walk on a large network, the cover time of a large target set was shown to obey a universal probability distribution of Gumbel form \cite{chupeau2015}. Notably, the only timescale in this limiting cover time probability distribution is the so-called global mean FPT, which is the mean FPT to a given target node averaged over random walks starting from all nodes in the network \cite{chupeau2015}.

The purpose of this paper is to study cover times of many searchers. Specifically, let $\{X_n\}_{n=1}^N$ be $N\ge1$ independent and identically distributed (iid) realizations of a random walk $X=\{X(t)\}_{t\ge0}$. If $S_N=\{S_N(t)\}_{t\ge0}$ denotes the set of nodes visited by at least one of the $N$ searchers by time $t\ge0$,
\begin{align}\label{eq:SN}
    S_N(t)
    :=\cup_{n=1}^N\cup_{s=0}^t X_n(s)\subseteq I,
\end{align}
then the cover time of $U_{\text{T}}\subseteq I$ of these multiple searchers is
\begin{align}\label{eq:sigmaN}
    \sigma_N
    :=\inf\{t>0:U_{\text{T}}\subseteq S_N(t)\}.
\end{align}
In words, $\sigma_N$ is the time needed for every node in the target to be visited by at least one searcher. Prior works have argued that the cover time for $N$ searchers can be obtained by simply rescaling the cover time of a single searcher \cite{chupeau2015, dong2023}. In particular, it was shown that if $N$ is not too large, then
\begin{align}\label{eq:previous}
    \sigma_N
    \approx \sigma/N.
\end{align}

\begin{figure}
    \centering
    \includegraphics[width=.9\linewidth]{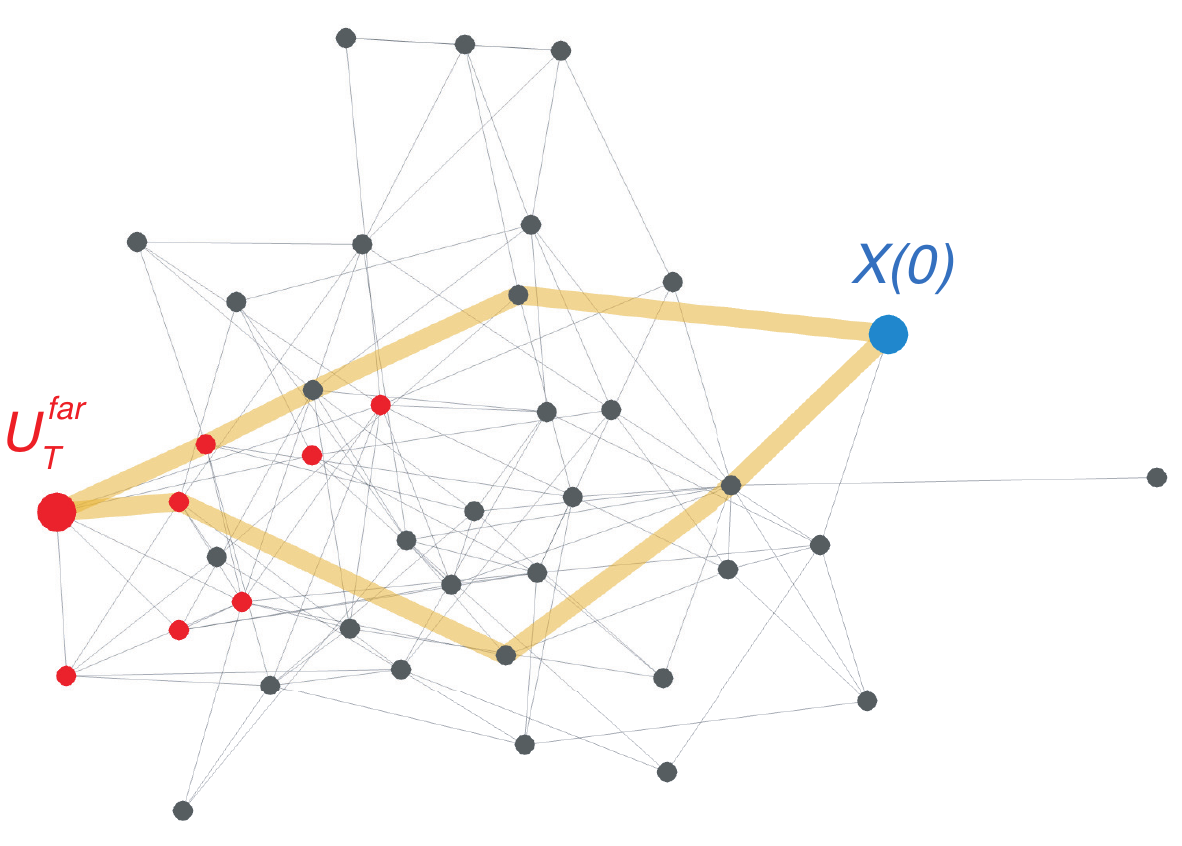}
    \caption{Example network. The searchers start at $X(0)$ (blue node), the red nodes are the target $U_{\text{T}}$, and the yellow curves depict the shortest paths a searcher can travel to cover the farthest part of the target (large red node labeled $U_\text{T}^\text{far}$).}
    \label{fig1}
\end{figure}

In this paper, we obtain an explicit formula for all the moments of the many searcher cover time. Specifically, we prove that for any moment $m\in(0,\infty)$,
\begin{align}\label{eq:main}
    \E[\sigma_N^m]
    \sim\frac{K_m}{N^{m/d}}\quad\text{as }N\to\infty,
\end{align}
where $d\ge1$ is the smallest number of steps a searcher must take to reach the farthest part of the target and the constant $K_m$ depends on the jump rates of the network along these geodesic paths to the farthest part of the target ($K_m$ is given explicitly in Theorem~\ref{thm:main}, see Figure~\ref{fig1} for an illustration). Throughout this paper, $f\sim g$ denotes $f/g\to1$. We prove \eqref{eq:main} in the general setting that the searchers are continuous-time Markov chains on a finite or countably infinite state space and the target $U_{\text{T}}$ consists of finitely many states. We make only mild assumptions on the network jump rates and the searcher initial position probability distribution in order to avoid trivial cases in which $\sigma_N=0$ or $\sigma_N=\infty$ (see Theorem~\ref{thm:main} for the precise statement). 

We make three comments on \eqref{eq:main}. First, comparing \eqref{eq:previous} with \eqref{eq:main} shows that the decay of $\sigma_N$ slows down markedly from $N^{-1}$ to $N^{-1/d}$ for $N$ sufficiently large (these decay rates are identical if and only if $d=1$, which is the simple case that searchers need only take a single step to cover the target).

Second, the integer $d\ge1$ and the constant $K_m$ in \eqref{eq:main} depend only on the jump rates along the shortest path(s) from the starting location(s) to the farthest part(s) of the target. In particular, changes to the network outside these geodesic path(s) do not affect $\sigma_N$ for large $N$. This is in stark contrast to the cover time of a single searcher, $\sigma$, which depends on the entire network \cite{chupeau2015}. The reason for this discrepancy is that a single searcher (or a few searchers) may wander around the network before covering the target, whereas cover times for many searchers are determined by searchers which take a direct path to cover the target.

Third, only the farthest parts of the target affect the integer $d\ge1$ and the constant $K_m$ in \eqref{eq:main}. That is, the many searcher cover time is unaffected by adding nodes or deleting nodes from the target which are strictly closer than the farthest parts of the target. For example, the cover time of the entire network (i.e.\ $U_{\text{T}}=I$) becomes identical to the cover time of only the nodes which are farthest from the initial searchers positions as $N\to\infty$.

The rest of the paper is organized as follows. In Section~\ref{sec:main}, we prove \eqref{eq:main} and give  $d$ and $K_m$ explicitly. In section~\ref{sec:lattice}, we analyze $K_m$ in the case of a simple random walk on a periodic lattice of arbitrary dimension. In section~\ref{sec:numerics}, we illustrate our results with stochastic simulations. We conclude by discussing relations to prior work and possible extensions. {In particular, we compare the results in the present paper to our recent results for cover times of many diffusive (or subdiffusive) searchers on a continuous state space \cite{kim2023}.}

\section{General theory}\label{sec:main}

We now present the general theory for the cover time for many Markovian searchers on a discrete state space.

\subsection{Inclusion-exclusion}

Let $X=\{X(t)\}_{t\ge0}$ be a continuous-time Markov chain on a finite or countably infinite state space $I$. The dynamics of $X$ are encoded in its infinitesimal generator matrix $Q=\{q(i,j)\}_{i,j\in I}$ \cite{norris1998}. The off-diagonal entries of $Q$ (i.e.\ $q(i,j)\ge0$ for $i\neq j$) are nonnegative and give the rate that $X$ jumps from $i\in I$ to $j\in I$. The diagonal entries of $Q$ are nonpositive (i.e.\ $q(i,i)\le0$ for all $i\in I$) and are chosen so that $Q$ has zero row sums. Assume that $\sup_{i\in I}|q(i,i)|<\infty$ so that $X$ cannot take infinitely many jumps in finite time.

Let $\{X_{n}(t)\}_{n=1}^N$ be $N$ iid realizations of $X$. For each subset of states $J\subset I$, let $\tau_n(J)$ denote the FPT of $X_n$ to $J$,
\begin{align*}
\tau_{n}(J)
:=\inf\{t>0:X_{n}(t)\in J\},
\end{align*}
and define the corresponding fastest FPT,
\begin{align}\label{eq:efpts}
T_{N}(J)
:=\min\{\tau_{1}(J),\dots,\tau_{N}(J)\}.
\end{align}

Let $S_N(t)$ denote the set of states visited by at least one of the searchers by time $t\ge0$ as in \eqref{eq:SN}. For a finite set of target states ${{U}}_{\text{T}}\subseteq I$, define the cover time $\sigma_N$ of these $N$ searchers as in \eqref{eq:sigmaN}. 

By the inclusion-exclusion principle \cite{durrett2019},
\begin{align}\label{eq:inclusionexclusion}
\begin{split}
    &\P(\sigma_{N}>t)
=\P(\cup_{i\in {{U}}_{\text{T}}}\{T_{N}(i)>t\})\\
&=\sum_{i=1}^{|U_{\text{T}}|} \Big((-1)^{i-1}\sum_{J\subseteq U_{\text{T}}, |J|=i} \P(\cap_{j\in J}\{T_{N}(j)>t\})\Big)\\
&=\sum_{i=1}^{|U_{\text{T}}|} \Big((-1)^{i-1}\sum_{J\subseteq U_{\text{T}}, |J|=i} \P(T_{N}(J)>t)\Big),
\end{split}
\end{align}
where the inner sum is over all subsets $J$ of the target $U_{\text{T}}$ of size $|J|=i$. 
Now, the mean of any nonnegative random variable $Z\ge0$ is given by the integral of its survival probability \cite{durrett2019},
\begin{align*}
    \E[Z]
    =\int_0^\infty \P(Z>z)\,\dd z.
\end{align*}
Therefore, setting $t=s^{1/m}$ for $m\in(0,\infty)$ and integrating \eqref{eq:inclusionexclusion} from $s=0$ to $s=\infty$ yields the following representation for the $m$th moment of the cover time $\sigma_N$ in terms of the $m$th moments of the fastest FPTs in \eqref{eq:efpts},
\begin{align}\label{eq:coverextreme}
\E[\sigma_{N}^{m}]
&=\sum_{i=1}^{|U_{\text{T}}|} \Big((-1)^{i-1}\sum_{J\subseteq U_{\text{T}}, |J|=i} \E[(T_{N}(J))^{m}]\Big).
\end{align}

\subsection{Fastest FPTs}

In light of \eqref{eq:coverextreme}, we now study moments of $T_{N}(J)$ for $J\subset U_{\text{T}}$ in order to study moments of $\sigma_N$. Let $\rho$ denote the initial distribution of $X$,
\begin{align*}
\rho
=\{\rho(i)\}_{i\in I}
=\{\P(X(0)=i)\}_{i\in I},
\end{align*}
and define the support of $\rho$,
\begin{align*}
\text{supp}(\rho)
:=\{i\in I:\rho(i)>0\}.
\end{align*}

To avoid trivial cases, we make the following two assumptions. First, assume that the target is not entirely contained in the support of the initial distribution,
\begin{align}\label{eq:startin}
    U_{\text{T}}\not\subset\text{supp}(\rho).
\end{align}
Second, assume that for each state $j\subset U_{\text{T}}$,
\begin{align}\label{eq:finite}
    \E[T_N(j)]<\infty\quad\text{for some }N\ge1.
\end{align}
If \eqref{eq:startin} is violated, then the searchers can cover the target just by their initial placement (i.e.\ without moving) and the problem is trivial. If \eqref{eq:finite} is violated, then $\E[\sigma_N]=\infty$ for all $N\ge1$ and the problem is trivial. Note that \eqref{eq:finite} is assured to hold if $X$ is irreducible on a finite state space $I$.

If $J\cap\text{supp}(\rho)\neq\varnothing$, then $\E[T_{N}(J)]$ vanishes exponentially fast as $N\to\infty$. If $J\cap\text{supp}(\rho)=\varnothing$, then Theorem~1 in \cite{lawley2020networks} implies 
\begin{align}\label{eq:ffpt}
\E[(T_{N}(J))^{m}]
\sim\frac{\Gamma(1+m/d(J))}{(A(J)N)^{m/d(J)}}\quad\text{as }N\to\infty,
\end{align}
where $d(J)\ge1$ is the smallest number of jumps that $X$ must make to reach $J$ from $\text{supp}(\rho)$ and $A(J)>0$ involves the jump rates along such shortest paths. We now define $d(J)$ and $A(J)$ more precisely.

\subsection{Defining $d(J)$ and $A(J)$}

Following \cite{lawley2020networks}, define a path $\PP$ of length $d\in\mathbb{Z}_{\ge0}$ from $i_{0}\in I$ to $i_{d}\in I$ to be a sequence of $d+1$ states in $I$,
\begin{align}\label{path}
\PP
=(\PP(0),\dots,\PP(d))
=(i_{0},i_{1},\dots,i_{d})\in I^{d+1},
\end{align}
so that
\begin{align}\label{explain}
q(\PP(j),\PP(j+1))
>0,\quad\text{for }j\in\{0,1,\dots,d-1\},
\end{align}
where $Q=\{q(i,j)\}_{i,j\in I}$ is the generator of $X$. 
The condition in \eqref{explain} ensures that $X$ has a strictly positive probability of following the path $\PP$.

For a path $\PP\in I^{d+1}$, let $\lambda(\PP)$ be the product of the rates along the path,
\begin{align}\label{eq:lambda}
\lambda(\PP)
:=\prod_{i=0}^{d-1}q(\PP(i),\PP(i+1))>0.
\end{align}
Let $d_{\min}(I_{0},I_{1})\in\mathbb{Z}_{\ge0}$ denote the length of the shortest path from $I_{0}\subset I$ to $I_{1}\subset I$,
\begin{align}\label{eq:dmin}
d_{\min}(I_{0},I_{1})
:=
\inf\{d:\PP\in I^{d+1},\PP(0)\in I_{0},\PP(d)\in I_{1}\}.
\end{align}
In words, $d_{\min}(I_{0},I_{1})$ is the smallest number of jumps that $X$ must take to go from $I_{0}$ to $I_{1}$. 
Define the set of all paths from $I_{0}$ to $I_{1}$ with this minimum length $d_{\min}(I_{0},I_{1})$,
\begin{align}\label{S}
\begin{split}
&\mathcal{S}(I_{0},I_{1})\\
&:=\{\PP\in I^{d+1}:\PP(0)\in I_{0},\PP(d)\in I_{1},d=d_{\min}(I_{0},I_{1})\}.
\end{split}
\end{align}
Define
\begin{align}\label{eq:Lambdathm}
\Lambda(\rho,I_{1})
:=\sum_{\PP\in\mathcal{S}(\textup{supp}(\rho),I_{1})}\rho(\PP(0))\lambda(\PP).
\end{align}
To unpack the meaning of $\Lambda(\rho,I_{1})$ in words, first consider the case that $\rho(i_{0})=1$ for some $i_{0}\in I$, which means that $X(0)=i_0$ with probability one and $\text{supp}(\rho)=i_{0}$. If there is only one path with the minimum number of jumps $d_{\min}(i_{0},I_{1})$, then $\Lambda(\rho,I_{1})$ is merely the product of the jump rates along this path (i.e.\ $\lambda(\PP)$ in \eqref{eq:lambda}). If there is more than one shortest path, then $\Lambda(\rho,I_{1})$ is the sum of the products of the jump rates along these paths. Lastly, if $\rho(i)\neq1$ for all $i\in I$ (meaning the initial searcher distribution is over multiple nodes), then $\Lambda(\rho,I_{1})$ sums the products of the jump rates along all the shortest paths, where the sum is weighted according to $\rho$.

Now that we have defined $d_{\min}$ in \eqref{eq:dmin} and $\Lambda$ in \eqref{eq:Lambdathm}, we define $d(j)$ and $A(J)$ in \eqref{eq:ffpt} via
\begin{align*}
d(J)
&=d_{\min}(\text{supp}(\rho),J)\ge1,\\
A(J)
&=\frac{\Lambda(\rho,J)}{(d(J))!}>0.
\end{align*}

\subsection{Cover times}

Having determined the large $N$ behavior of the moments of $T_N(J)$ in \eqref{eq:ffpt}, we now determine the large $N$ behavior of the moments of $\sigma_N$ via \eqref{eq:coverextreme}. 
Define
\begin{align*}
d
:=\sup_{j\in{{U}}_{\text{T}}}d_{\min}(\text{supp}(\rho),j)\ge1,
\end{align*}
which is the smallest number of jumps required to reach the farthest part of the target. Further, define the set of nodes in the target set which are distance $d\ge1$ from the initial distribution,
\begin{align}\label{eq:UTfar}
    U_{\text{T}}^{\text{far}}
    :=\{j\in U_{\text{T}}:d_{\min}(\text{supp}(\rho),j)=d\}
    \subseteq U_{\text{T}}\subseteq I.
\end{align}

Hence, \eqref{eq:coverextreme} and \eqref{eq:ffpt} imply
\begin{align*}
&\lim_{N\to\infty}N^{m/d}\E[\sigma_{N}^{m}]\\
&=\sum_{i=1}^{|U_{\text{T}}|} \Big((-1)^{i-1}\sum_{J\subseteq U_{\text{T}},|J|=i} \lim_{N\to\infty}N^{m/d}\E[(T_{N}(J))^{m}]\Big)\\
&=\sum_{i=1}^{|U_{\text{T}}^{\text{far}}|} \Big((-1)^{i-1}\sum_{J\subseteq U_{\text{T}}^{\text{far}}, |J|=i} \lim_{N\to\infty}N^{m/d}\E[(T_{N}(J))^{m}]\Big)\\
&=K_m,
\end{align*}
where
\begin{align}\label{eq:Km}
    K_m
    :=\Gamma\big(1+\frac{m}{d}\big)\sum_{i=1}^{|U_{\text{T}}^{\text{far}}|} \bigg[(-1)^{i-1}\sum_{J\subseteq U_{\text{T}}^{\text{far}}, |J|=i} \Big(\frac{d!}{\Lambda(\rho,J)}\Big)^{\frac{m}{d}}\bigg],
\end{align}
where the inner sum is over all subsets $J$ of $U_{\text{T}}^{\text{far}}$ in \eqref{eq:UTfar} of size $|J|=i$.

In summary, we have proven the following theorem.

\begin{theorem}\label{thm:main}
    Under assumptions \eqref{eq:startin} and \eqref{eq:finite}, we have that for any moment $m\in(0,\infty)$,
    \begin{align*}
        \E[\sigma_N^m]
        \sim \frac{K_m}{N^{m/d}}\quad\text{as }N\to\infty.
    \end{align*}
\end{theorem}

Theorem~\ref{thm:main} shows that the simple scaling in \eqref{eq:previous} breaks down for large $N$. Indeed, the large $N$ decay of $\sigma_N$ is in general much slower than predicted by \eqref{eq:previous} except for the case that $d=1$. Further, while the cover time of a single searcher depends on the entire network, Theorem~\ref{thm:main} shows that the cover time of many searchers only depends on network properties along the shortest path(s) to the farthest part(s) of the target. In particular, the many searcher cover time is unaffected by any changes to the network away from these geodesic paths. In addition, the many searcher cover time only depends on the farthest parts of the target. Put another way, the many searcher cover time is unaffected by adding nodes or deleting nodes from the target which are strictly closer than the farthest parts of the target. For example, the cover time of the entire network (i.e.\ $U_{\text{T}}=I$) becomes identical to the cover time of only the nodes which are farthest from the initial searcher positions as $N\to\infty$. We illustrate these points with stochastic simulations in section~\ref{sec:numerics}.

\section{Periodic lattice in arbitrary dimension}\label{sec:lattice}

Theorem~\ref{thm:main} is a very general result that holds for Markovian searchers on general discrete state spaces (i.e.\ random walkers on networks of discrete nodes). Examining the formula in \eqref{eq:Km} shows that while the constant prefactor $K_m$ depends only on the network jump rates along the shortest path(s) from the searcher initial position(s) to the farthest part(s) of the target, this dependence is far from trivial. In this section, we investigate $K_m$ in the case that the searchers are simple random walks on periodic lattices, which is a case that has received much attention for a single searcher \cite{nemirovsky1990, yokoi1990, brummelhuis1991, hemmer1998, mendoncca2011, grassberger2017, dembo2004, ding2012}. We first suppose that the lattice is two dimensional and then consider a lattice of arbitrary dimension.

Consider a two dimensional square lattice $I$ with periodic boundaries (i.e.\ a lattice wrapped around a two-dimensional torus). Suppose all the searchers start at a single node $i_0\in I$,
\begin{align*}
    \P(X(0)=i_0)=\rho(i_0)=1.
\end{align*}
Suppose further that the number of nodes along each side of the lattice is even and given by $2l\ge2$ (and hence the total number of nodes is $|I|=(2l)^2\ge4$). It follows that there is a unique farthest node from the starting position $i_0\in I$ which we denote by $i_1\in I$ and is
\begin{align*}
    d=d_{\min}(i_0,i_1)=2l
\end{align*}
jumps away from $i_0$. Suppose the target $U_{\text{T}}\subseteq I$ is any set containing $i_1$, and thus $U_{\text{T}}^{\text{far}}$ in \eqref{eq:UTfar} is $i_1$,
\begin{align*}
  U_{\text{T}}^{\text{far}}
  =i_1
  \in U_{\text{T}}\subseteq I.
\end{align*}

If all the jumps in the lattice have the same rate ${r}>0$, then $\Lambda$ in \eqref{eq:Lambdathm} is given by
\begin{equation*}
    \Lambda(i_0,i_1) = {r}^{2l} |\mathcal{S}(i_0,i_1)|,
\end{equation*}
where $|\mathcal{S}(i_0,i_1)|$ is the number of paths from $i_0$ to $i_1$ with the minimum length $d=d_{\min}(i_0,i_1)=2l$. Since there is only one farthest target node (i.e.\ $U_{\text{T}}^{\text{far}}=i_1$), the constant $K_m$ in \eqref{eq:Km} takes the form
\begin{equation}\label{Cm_1}
    K_m = \frac{\Gamma(1+m/(2l))}{{r}^m}\left[\frac{(2l)!}{|\mathcal{S}(i_0,i_1)|}\right]^{m/(2l)}.
\end{equation}
The number of shortest paths from $i_0$ to $i_1$ is
\begin{equation}\label{eq:numpaths}
    |\mathcal{S}(i_0,i_1)| = 4{2l \choose l}.
\end{equation}
To see this, notice first that every path will have either all steps up and right, up and left, down and right, or down and left (and so four choices). Further, having made one of these four choices, say up and right, there will be exactly $l$ steps up and $l$ steps right, and thus there are ${2l \choose l}=(2l)!/(l!)^2$ many ways to choose when the $l$ steps up will be taken.

Therefore, \eqref{Cm_1} and \eqref{eq:numpaths} imply
\begin{align*}
    K_m = \frac{\Gamma(1+m/(2l))}{{r}^m}(l!/2)^{m/l}.
\end{align*}
Using Stirling's formula \cite{bender2013} yields the approximation
\begin{align*}
    K_m\sim \Big(\frac{l}{e{r}}\Big)^m\quad\text{as }l\to\infty.
\end{align*}

We now generalize this calculation to an $h$-dimensional periodic lattice for $h\ge1$. If the number of nodes along each of the $h\ge1$ sides is even and given by $2l\ge2$, then there is a unique farthest node $i_1\in I$ from the starting position $i_0\in I$, and the minimum distance from $i_0$ to $i_1$ is
\begin{align*}
d=d_{\min}(i_0,i_1)=hl    
\end{align*}
jumps. Hence, if the target $U_{\text{T}}$ contains $i_1$, then \eqref{eq:Km} becomes
\begin{equation}\label{eq:h}
    K_m = \frac{\Gamma(1+m/(hl))}{{r}^m}\left[\frac{(hl)!}{|\mathcal{S}(i_0,i_1)|}\right]^{m/(hl)},
\end{equation}
and \eqref{eq:numpaths} generalizes to the following multinomial formula,
\begin{equation}\label{eq:numpaths2}
    |\mathcal{S}(i_0,i_1)|
    = 2^h{hl \choose l, l, \ldots, l}
    = 2^h\frac{(hl)!}{(l!)^h}.
\end{equation}
Hence, \eqref{eq:h} and \eqref{eq:numpaths2} imply
\begin{align*}
    K_m = \frac{\Gamma(1+m/(hl))}{{r}^m}(l!/2)^{m/l}.
\end{align*}
Using Stirling's formula \cite{bender2013} yields the approximation
\begin{align*}
    K_m\sim \Big(\frac{l}{e{r}}\Big)^m\quad\text{as }l\to\infty.
\end{align*}

\section{Numerical simulations} \label{sec:numerics}

In this section, we compare the results of our analysis to numerical simulations on two types of networks: a square lattice with periodic boundaries and a randomly constructed network.

\subsection{Periodic lattice in two dimensions}

\begin{figure*}
    \centering
    \includegraphics[width = .8\linewidth]{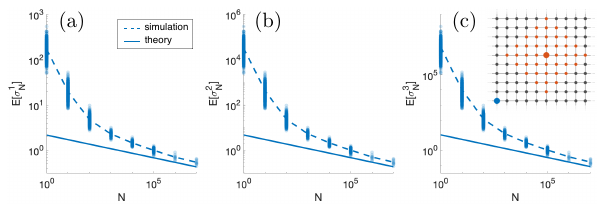}
    \caption{Comparison of stochastic simulations (\textit{dashed curve}) on the lattice network (inset in panel (c)) and the moment formula (\textit{solid curve}) in Theorem~\ref{thm:main} for the $m=1$ moment (panel (a)), $m=2$ moment (panel (b)), and $m = 3$ moment (panel (c)). The circle markers at $N = 10^0,\cdots,10^6$ are scatter plots of individual stochastic realizations of the cover time $\sigma_N$. In the inset in panel (c), the blue node is the searcher starting location, the red nodes are the target, and the big red node is the farthest node from the starting location.}
    \label{figlatticeA}
\end{figure*}

\begin{figure}
    \centering
    \includegraphics[width = .8\linewidth]{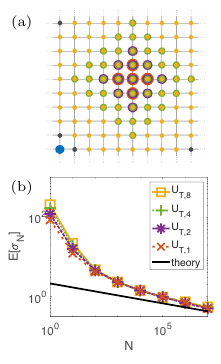}
    \caption{The cover time is determined by the farthest target nodes. Panel (a) illustrates the starting point of the searchers (\textit{blue}) and the nested target sets $U_{\text{T},k}$ in \eqref{eq:nested} for $k = 1$ (\textit{red}), $k = 2$ (\textit{purple}), $k = 4$ (\textit{green}), and $k = 8$ (\textit{yellow}). Note that $U_\text{T}^\text{far} \subset U_{\text{T},1} \subset U_{\text{T},2} \subset U_{\text{T},4} \subset U_{\text{T},8}$.
    Panel (b) compares stochastic simulations and the theoretical result of the mean cover time for various choices of the target set. Network parameters are the same as Figure~\ref{figrandomA}.}
    \label{figlatticeB}
\end{figure}

We now present results from stochastic simulations of a simple random walk on a two-dimensional square lattice as in section~\ref{sec:lattice} with $2l=10$ nodes on each side (i.e.\ $|I|=(2l)^2=100$ total vertices). 

To simulate $N\ge1$ searchers (i.e.\ $N$ iid realizations of $X$), we use the Gillespie algorithm on the Markov chain $(X_1(t),\dots,X_N(t))_{t\ge0}$ on the state space $I^N$ \cite{gillespie1976}. Specifically, let $r(i,j)$ be the ``reaction rate'' from $i$ to state $j$ with $i \neq j$ (because there is no ``jump'' to the same vertex). Since the jump rate of a single searcher is $q(i,j)$, the reaction rate is proportional to the number of searchers at state $i$:
\begin{equation*}
    r(i,j) = q(i,j) c_i,
\end{equation*}
where $c_i$ is the number of searchers at state $i$. To numerically compute the moments of the cover time $\sigma_N$, we use the Monte Carlo method with $M\gg1$ realizations of the cover time of $N$ searchers. We choose $M = 10^3$ when $N \leq 10^4$ and $M = 10^2$ when $N > 10^4$ because the variance of $\sigma_N$ decreases in $N$.

We plot the results of these stochastic simulations in Figure~\ref{figlatticeA}. This figure shows that the first, second, and third moments of the cover time $\sigma_N$ approach the theoretical prediction of Theorem~\ref{thm:main} for large $N$. The inset in panel (c) of Figure~\ref{figlatticeA} shows the lattice network, where all searchers start from the large blue node, the red nodes denote the target set $U_{\text{T}}$, and the large red node denotes the farthest part of the target (i.e.\ $U_{\text{T}}^{\text{far}}$ in \eqref{eq:UTfar}).

In Figure~\ref{figlatticeB}, we plot the results of stochastic simulations on this same network but with the following nested target sets,
\begin{equation}\label{eq:nested}
    U_{\text{T},k} = \{i \in I : d_{\min}(i,U_\text{T}^\text{far}) \leq k \}, \quad k\in\{1,2,4,8\},
\end{equation}
where
\begin{equation*}
U_\text{T}^\text{far} = U_{\text{T},0}, \quad U_{\text{T},k} \subseteq U_{\text{T},k+1}.    
\end{equation*}
That is, the the different curves in Figure~\ref{figlatticeB} show the different numerically computed cover times $\sigma_N$ for when the target set is either $U_{\text{T},8}$, $U_{\text{T},4}$, $U_{\text{T},2}$, or $U_{\text{T},1}$. Importantly, these four target sets share the same set $U_\text{T}^\text{far}$ in \eqref{eq:UTfar}. Therefore, Theorem~\ref{thm:main} implies that the cover times for these different target sets become identical for large $N$, which is indeed shown in Figure~\ref{figlatticeB}.

\subsection{Random network}

\begin{figure*}
    \centering
    \includegraphics[width=.8\linewidth]{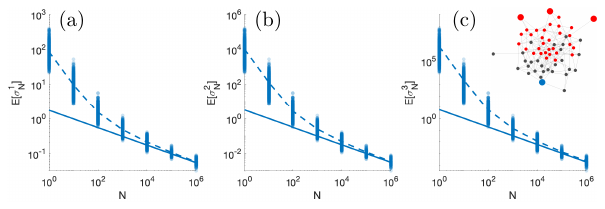}
    \caption{Comparison of stochastic simulations (\textit{dashed curve}) on the network (inset in panel (c)) and the moment formula (\textit{solid curve}) in Theorem~\ref{thm:main} for the $m=1$ moment (panel (a)), $m=2$ moment (panel (b)), and $m = 3$ moment (panel (c)). The circle markers at $N = 10^0,\cdots,10^6$ are scatter plots of individual stochastic realizations of the cover time $\sigma_N$. In the inset in panel (c), the blue node is the searcher starting location, the red nodes are the target, and the big red nodes are the farthest nodes from the starting location.}
    \label{figrandomA}
\end{figure*}

\begin{figure}
    \centering
    \includegraphics[width = .8\linewidth]{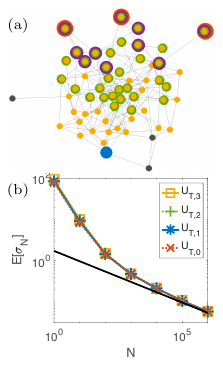}
    \caption{The cover time is determined by the farthest target nodes. Panel (a) illustrates the starting point of the searchers (\textit{blue}) and the nested target sets $U_{\text{T},k}$ in \eqref{eq:nested} for $k = 0$ (\textit{red}), $k = 1$ (\textit{purple}), $k = 2$ (\textit{green}), and $k = 3$ (\textit{yellow}). Note that $U_\text{T}^\text{far} = U_{\text{T},0} \subset U_{\text{T},1} \subset U_{\text{T},2} \subset U_{\text{T},3}$.
    Panel (b) compares stochastic simulations and the theoretical result of the mean cover time for various choices of the target set. Network parameters are the same as Figure~\ref{figrandomA}.}
    \label{figrandomB}
\end{figure}

We now perform stochastic simulations on a randomly constructed network. To create the continuous-time Markov chain for a single searcher, we first create a graph by randomly connecting $|I| \geq 1$ vertices by $E$ undirected edges (we take $|I|=60$ and $E=180$). We then assign jump rates to each undirected edge independently according to a uniform distribution. More precisely, if $Q = \{q(i,j)\}_{i,j \in I}$ denotes the infinitesimal generator matrix of the Markov chain, then the diagonal entries, $q(i,i) \leq 0$ are chosen so that $Q$ has zero row sums, and the off-diagonal entries, $q(i,j) \geq 0$ with $i \neq j$, are
\[
    q(i,j) = \begin{cases}
        U_{i,j} & \text{if there is an edge between } i \text{ and } j, \\
        0 & \text{otherwise},
    \end{cases}
\]
where $\{U_{i,j}\}_{i,j \in I}$ are independent uniform random variables on $[1/2, 3/2]$. We use the same Gillespie stochastic simulation method as above to obtain the cover time moments.

We plot the results of these stochastic simulations in Figure~\ref{figrandomA}. This figure shows that the first, second, and third moments of the cover time $\sigma_N$ approach the theoretical prediction of Theorem~\ref{thm:main} for large $N$. The inset in panel (c) of Figure~\ref{figrandomA} shows the topology of the randomly constructed network for the simulations, where all searchers start from the large blue node, the red nodes denote the target set $U_{\text{T}}$, and the large red nodes denote the farthest parts of the target (i.e.\ $U_{\text{T}}^{\text{far}}$ in \eqref{eq:UTfar}).

Analogous to Figure~\ref{figlatticeB}, we plot in Figure~\ref{figrandomB} the results of stochastic simulations on this randomly constructed network but with nested target sets as in \eqref{eq:nested} for $k\in\{0,1,2,3\}$. Since these four target sets share the same set $U_\text{T}^\text{far}$ in \eqref{eq:UTfar}, Theorem~\ref{thm:main} implies that the cover times for these different target sets become identical for large $N$, which is indeed shown in Figure~\ref{figrandomB}.

\section{Discussion}

In this paper, we studied the cover time $\sigma_N$ for $N\gg1$ random searchers on an arbitrary discrete network. We found an explicit formula for all the moments of $\sigma_N$ which depends only on network properties along the shortest paths from the searcher starting locations to the farthest parts of the target. These results contrast qualitatively with prior results for single searchers.

Many searcher cover times were recently studied for the case that the searchers move by diffusion (or subdiffusion) on a continuous state space \cite{majumdar2016, kim2023}. For searchers which move by a continuous drift-diffusion process with characteristic diffusivity $D>0$ in an arbitrary space dimension, it was proven in \cite{kim2023} that
\begin{align}\label{eq:diffusion}
    \E[\sigma_N^m]
    \sim\Big(\frac{L^2}{4D\ln N}\Big)^m\quad\text{as }N\to\infty,
\end{align}
where $L>0$ is a certain geodesic distance from the searcher starting locations to the farthest parts of the target. The result \eqref{eq:diffusion} extended a prior result of Majumdar, Sabhapandit, and Schehr \cite{majumdar2016} that was shown for Brownian motion in one space dimension. 

The result in \eqref{eq:diffusion} for searchers on a continuous state space bears some resemblance to the result in the present paper for searchers on a discrete state space. In both the continuous and discrete scenarios, $\sigma_N$ depends on the shortest distance to the farthest parts of the target, which can be understood by observing that $\sigma_N$ is determined by extreme searchers which take a direct path to cover the target. One notable difference between the two scenarios is that the discrete result in Theorem~\ref{thm:main} in the present paper depends in a rather complicated way on the number of shortest such paths, whereas the continuous state space result in \eqref{eq:diffusion} does not.

We point out that though $\sigma_N$ depends on extremely rare events if $N\gg1$, the analysis of $\sigma_N$ does not fit into the framework of classical extreme value theory. Extreme value theory is a branch of probability theory and statistics that studies the behavior of rare events \cite{haanbook}, with classical results giving the probability distribution of the minima (or maxima) of a large collection of iid random variables \cite{fisher1928}. Extreme value theory can thus be directly applied to analyze fastest FPTs \cite{lawley2023review, lawley2020dist, madrid2020comp, lawley2020sub, lawley2023super}. However, extreme value theory is not directly applicable to cover times since a cover time of multiple random searchers is not simply the minimum of a collection of iid random variables. In particular, realizations of $\sigma_N$ tend to describe multiple extreme searchers which separately cover different parts of the target. {We note that a very interesting related question about the maximum of non-iid visitation times has been considered in the context of so-called starving random walks \cite{regnier2023}.}

While we proved our results for searchers which are general Markovian random walkers, our assumption that the walkers are continuous-time Markov chains means that the times between walker jumps are necessarily exponentially distributed. One possible avenue for future work would be to relax this assumption of exponential times. In light of prior results on extreme FPTs for random walkers with non-exponential jump times (see Theorems~3 and 4 in \cite{lawley2020networks}), we expect similar results to that found in Theorem~\ref{thm:main}, with the differences being determined by the short-time behavior of the jump time distribution.


\begin{acknowledgments}
SDL was supported by the National Science Foundation (Grant Nos.\ DMS-2325258 and CAREER DMS-1944574). 
\end{acknowledgments}

\bibliography{library.bib}
\bibliographystyle{unsrt}

\end{document}